# Growth of GaAs nanowires on Au, Au / Pd, Ag, Ni, Ga, Cu, Al, Ti  metal films

G. Statkutė[1]*
*[1]Department of Micro and Nanosciences, Aalto University, School of Electrical Engineering.
Tietotie 3, 02150 Espoo, Finland*

E-mail: g.statkute@gmail.com

## Abstract

GaAs nanowires (nw's) were grown by metalorganic vapor phase epitaxy on evaporated metal films (Au, Au / Pd, Ag, Ni, Ga, Cu, Al, Ti). The samples were characterized by scanning electron microscope (SEM) and transmission electron microscope (TEM). SEM images reveal that nanowires grow directly on the metals. TEM characterization shows crystalline nanowire (nw) structure originating from Au. Article presents state of the art about nanowire-metal interface growth and enumerates nanowire contacting methods with metals.

## Introduction

Semiconductor nanowires are promising nanostructures for optoelectronics, electronics, memory devices, logical circuits, spintronics, optics, renewable energy and energy storage, chemistry, medicine. A part of device applications requires electrical contacting with metals. In this paper we present  results on GaAs nanowires grown directly on metal films by metalorganic vapor phase epitaxy (MOVPE). Several metals, Au, Au / Pd, Ag, Ni, Ga, Cu, Al, and Ti, were used. This new technology is interesting from several points of view: scientifically - epitaxial growth of semiconductor on metals, and industrially - direct junction formation between metal and nanowire.

Concerning fabrication of metal-semiconductor interface for nw's, a huge advancement was semiconductor growth around the metal, so called epitaxial overgrowth [1], which with additional lithography step was applied for the manufacturing of nw's in 2004 by *Ooike et al.* [2].

As far as we know, we were among the first to report semiconductor nanowire - metal layer epitaxy results about GaAs nanowires grown on metallic substrates in 2008 in conferences [3,4], and in ArXive [5]. Growth of other nw's semiconductors on metals has been reported independently (because authors does not cite others achievements) by manufacturing nanostructured metal-semiconductor hetero-junction nanowires, growing nanowires on metal layers or crystallites, evaporating metals on nw's, [6-27,87,88,90] and obtaining semiconductor nw's by metal oxidation [89]. Presented here results comparing to the state of the art shows the best quality images of III-V semiconductor nw growth on the metal layers and HRTEM characterization implies possibility of semiconductor nanowire epitaxy on the metal layer, which was not yet demonstrated as far as we found.

Besides nw growth on the metal, there is a range of metal-nanowire contacting methods [28-81], which are touched in the discussion. We show results about GaAs nanowires contacting from substrate side with metal during the MOVPE growth and discuss pros and cons comparing with other methods.

## Experimental methods

Various metals, Au, Au/Pd, Ag, Cu, Ni, Ga, Al and Ti with the thickness of 50-200 nm were evaporated on c–plane sapphire. The purity of the source metals was 99.99%. The metalized substrates were annealed *in situ* in an infrared heated horizontal flow MOVPE reactor (*Thomas Swan*) from 400 ºC to 750 ºC for several minutes before the growth. For a part of the growth runs (on Ag, Cu, Ni, Ga, Al, Ti) 8-16 sccm flow of trimethylgallium (TMG) was introduced to the reactor for 10-90 s at the temperature of 450-750 ºC to form Ga droplets needed for the nw growth. GaAs nw's were subsequently grown at 400-550 ºC for 3 min. During the growth using $H_2$ as the carrier gas the precursor flows were 8-16 sccm and 70-140 sccm for TMG and tertiarybutylarsine (TBA), respectively. All samples were characterised with scanning electron microscope (SEM) using secondary and back-scattered electron detectors. The GaAs nw's grown directly on Au without Ga droplets were further characterised with high-resolution transmission electron microscope (HRTEM) at accelerating voltage of 200 kV.

## Results and discussion

SEM characterization shows the formation of GaAs nanowires with the metal droplet on the top and in solid contact with metal substrate. The results of nw's on various metals are shown in Figs. 1-3. Only random up to 1-2 μm long single nanowires grew on Au, Ti, Al, Cu and Ga. In the case of Ti, Ni and Al, additionally, metallic surface is covered with thin semiconductor crystallites. Denser nanowire array is obtained on Ag. Very dense and long (up to 10 μm) nanowires were achieved on Ni. HRTEM of GaAs nanowires grown on Au shows firm contact between Au and GaAs nw, where it is possible to see crystalline GaAs lattice originating from the metal (Fig. 3), which implies epitaxy of semiconductor on a metal. Crystalline lattice is obviously distorted at the metal side. The GaAs nanowire starts growing as a crystallite, like foot, and later after some 10-20 nm of height abruptly achieves its final diameter. This foot can be also observed in nanowires grown on Cu (Fig. 1). Nanowires on Ag, Ga, Ti and Ni start growing as if from pedestal of height of 100-500nm (Figs. 1 & 2,). Although growth parameters were varied, the influence of Ga droplet size and presence / absence, precursors concentrations and temperature were not thoroughly investigated. Results of GaAs nw growth on the metals were achieved in all given conditions ranges: 400-550 ºC, 8-16 sccm of TMG and 70-140 sccm for tertiarybutylarsine (TBA).

Speaking about the sate of the art, i narrow semiconductor nanowire growth on the metal to the case where semiconductor nanowire elemental compounds and metal are different, excluding well known and good reported cases where the nanowire is obtained by the metal oxidation, for instance see Ref. 89&11, and including epitaxial GaAs nw's growth on Ga. The possibility to achieve well investigated and promising metallic contacts from Au, Cu, Ti, Al, Ni, Ag to the III-V semiconductor nw's bottom via epitaxial growth is a new branch of epitaxy promising a lot of applications.

In the selected state of the art there were only several attempts to grow/draw nw's on the metal films, but they do not show metal-semiconductor interface. *Greyson* et al. [10] in 2004 grew ZnO nw's on 40-200 nm Au dots, but he simply stated that nw grow on the Au drawing an image, not providing any high resolution image of ZnO nw's with Au interface. It is possible to argue that ZnO nw's grew via VLS mechanism from patterned Au islands with epitaxial contact with sapphire substrate. *Kim* at al. [87] in 2005 obtained Si nw's growth using Ni evaporated layer, but their SEM characterisation does not show Si-Ni interface and does not imply that Si nw's grew on the Ni, mostly they show that thin Si layer and Si nanowires form on the silica, the latter most probably consuming deposited Ni to the top metal seeds. In 2006 *Choi* et al. [14] published results about ZnO nw's growth on Au, Pt and Ti electrodes. Highest resolution image is with 5um scale bar, thus it is impossible to see metal - ZnO interface, nevertheless deciding by deposited metal thickness it is possible to conclude that ZnO grow on the metal, but nobody can state that they grow from the metal and have contact with it, nw's may be self-assembled or grow on self-assembled ZnO film. *Kim* et al. [21] in 2008 draw a model where ZnO nw's grew on the Au seeds, but they did not show any SEM or HRTEM image with this result. In 2009 *Chou* et al. published results about ZnO nw's growth on Cu electroplated substrate [18] and *Zalamai* et al. described ZnO nanorods on stainless steel [19]. *Chou* obtained thin film growth on the Cu which he did not investigated and did not show the metal - semiconductor interface, he also found that ZnO nw grew on the thin ZnO film but not on the Cu [18]. *Zalamai* does not investigated how ZnO grew on the steal and also did not show the interface between the metal and semiconductor leaving open questions if semiconductor layer or nanowires self assembly on the metal or if there is some junction between them [19]. In 2011 *Ye* et al. [22] described ZnO nw's growing horizontally from Au contacts. *Ye* showed low magnification images of ZnO on 5nm Cr + 50nm Au neither investigating if nw or thin film forms on the metal, nor metal-semiconductor interface possibility. *Benjamin* et al. [88] in 2014 show Ge and Si nw on Ag, Al, Au, Cr, Cu and Ni films with 20 um – 100 nm resolution, thus it is also not possible to see if there is a self-assembly on the metal layer or a contact with the metal. In my case, i present high magnification SEM images of nanowires interfaces with the metals and HRTEM of GaAs nanowire on Au.

Additionally, in the selected state of the art there were shown metal-semiconductor nanostructures heterojunctions: *Luo* et al. [6] in 2002 synthesized Si-Ni nw's heterojunctions, *Peña* et al. [7] electrodeposited metal-CuSe-metal in AAO templates, in 2003 *Sokolov* et al. [9] manufactured magnetoresistive Ni/NiO/Co nw's junctions, in 2006 *Luo* et al. [15] Ag-Si nw's heterojunction, in 2007 *He* et al. [17] - Si-Ni nanorods nanosprings. In 2013 *Chiu Chung-Hua* grew copper silicide/silicon nanowire heter nw's es in situ in TEM [23], *Haldar* et al. [24] grew epitaxially CdSe on selected facets of metal Au seeds and designed different shape hetero-nanostructures (flower, tetrapod, and core/shell). In 2015 *Polat* et al. [26] showed multilayered Cu/Si nanorods, *Bose* et al. in 2014 grew and modeled Au-ZnSe epitaxial heterojunction [90]. One more kind of heterojunctions includes nanotubes filled with metal and possibly semiconductor. For instance, *Hu* et al. [8] in 2003 showed Gallium-Filled Gallium Oxide–Zinc Oxide Composite Coaxial Nanotubes, in 2005 *Zhan* et al. [12] demonstrated In and Si heterojunction and *Hu* et al. [13] ZnS and Ga heterojunction in silica nanotube. In 2007 *Hu* et al. manufactured $Mg_3N_2$ and Ga heterojunction inside graphitic carbon nanotube [16]. Proposed application is a thermocouple. Third way to achieve nanostructured interface is to functionalise nanowires with metal crystallites, which found applications in surface enhanced Raman spectroscopy (SERS) [93], sensors [94], solar cells [95], photoelectrodes [96] via catalytic, plasmonic and Schottky field effects improving these devices performance. The forth group of nanostructured interfaces encompass core-shell metal-semiconductor nw's, mostly applicable in plasmonics, which were reviewed by *Jiang* et al. [25] in 2014. Core shell metal-semiconductor nw`s are also promising in invisible cloaking [91], lasing [91], SERS [97], AlAs core /Al shell (superconductor) epitaxial abrupt interface by *Krogstrup* et al. [27] in 2015 has promising applications on quantum computing. These nanostructured interfaces are not comparabe technologically with nw's growth on the metal layer, because these nw's require additional metalisation step each with 3D – 1D methods enumerated below in order to obtain bottom metallic contact. In my case, the bottom metal contact is manufactured instantly to millions of nw's. The metal layer epitaxy on the semiconductor scientifically is different from semiconductor nw's growth on the metal layer by epitaxial technique, because the former is achieved by evaporation and assembly with thermal annealing, while the latter proceeds via VLS or VSS mechanism. Nanowire epitaxy/assembly on the metal layer with clear interface was not shown in high-magnification images as far as we found.

There is a kind of semiconductor – metal epitaxy, where several semiconductor monolayers are assembled on the metal, for instance, in 2010 *Ronci* et al. [20] manufactured horizontal 2-4 atomic monolayers thickness Si nw's on Ag(110) surface and characterized with STM finding metallic conductivity. Authors does not propose applications except fundamental science investigation of nanoscale Si conductivity. This kind of epitaxy and kind of nw's differs a lot despite from the same name from thick vertical nanowire growth on the metal layers which it is presented here.

Here we have proposed a way to manufacture metal – nw semiconductor bottom-side contact during MOCVD growth: the quality of contact equals and exceeds 1D contact, while the density of contacted nw's equals to the case of array contact described in introduction. Our way of contacting has many advantages: (1) it acts as a substrate side / bottom metallic contact, (2) it has a smooth interface with nanowire and the same diameter as the nw, (3) it takes only few minutes to grow millions of contacted nanowires (or several hours if calculating metal evaporation and sample transfer time) thus it may be applicable in industrial scale, (4) it is interesting from scientific point of view as epitaxy on metals may introduce novel high efficiency plasmonic effects and novel electrical properties.

Various ways for metallic contacts to nanowires have been reported: contacting nw arrays, contacting single nanowire with bulk metallic contact (3D contact), contacting single nanowire with point contact (1D contact).

For example, array contact can be achieved by growing nw's on conducting substrate, than spin coating vertical nanowire arrays with insulator like PMMA [28] or silsesquioxanes SOG-400F [29] and evaporating metallic contacts on the top. As far as we know, nw array contacts demonstrated till now were achieved by growing nw on conducting non-metallic substrate, then spin-coating vertical nanowire arrays by an insulator and finally evaporating metallic contacts on the top. However, it is impossible to manufacture metal contacts on both sides of the nw array this way. Our approach enables a novel way to achieve bottom metallic contact.

3D contact to single nanowire can be prepared using: (1) electrical beam lithography (EBL) [30-33], (2) optical lithography [34], (3) in situ electrochemical technique [35,36], (4) selective area growth [37], (5) specially designed probe tips, which can manipulate and electrically contact nanowires [38,39] (6) dialectrophoretic manipulation [40-43], (7) Langmuir–Blodgett (L–B) technique and optical lithography [44], (8) optical tweezers for nanowire positioning (and fusing to the contacts) [45-46], (9) magnetic field force to align metal sputtered nanowire [47]. By 3D methods bulky compared to the diameter of the nanostructure, not matching in diameter contacts are achieved due lithography processes errors. In our case the contact has the same diameter as the nw. During nanostructure contacting with 3D methods such as in situ electrochemical technique, dielectrophoretic manipulation or magnetic field force to align metal sputtered nanowires the devices are lying horizontally which reduces their density. Moreover the contacts are flat, touching nanowire only from one side, while in our case we manufacture "volume" contacts to vertical nanowires. The last but not least disadvantage is that enumerated 3D methods deals only with single nanowire, thus are not efficient for industry, unlike the method presented in this article.

1D contact can be achieved by EBL with thermal annealing [72-81]; by scanning probe microscopy (SPM) [48-57], FIB [58-61], HRTEM with specially designed probes [62-64], SEM with specially designed probes [65-69] or can be obtained by spreading randomly metal and semiconductor nanowires [70]. Besides, 1D metal – nanowire contact, though not yet integrated in any kind of device, is always achieved in metal catalysed nanowires growth on the tip of nanowire -there are thousands or article publications - the first publication describing metal (Au, Pd, Pt, Ag, Cu, Ni) catalyzed silicon nanowire growth on silicon was published in 1964 by *Wagner* and *Ellis* [71]. Enumerated 1D methods are expensive and long lasting, their yield is 1 device per 30 min in the best cases, and therefore they are not applicable in industry. Typically, during nw contacting metallisation is carried out twice: first for catalytic metal nanostructure growth or lithography and later for contact deposition. The second step is long-lasting, additionally expensive lithographic contact patterning is involved. In our case fabrication of metallic contact is achieved in a single step and there is no need for the second metallisation. Speaking about the quality, only the contacts with subsequent thermal annealing step [72-81] has a similar quality as in our case, but they are manufactured only for 1 nw device per several hours of processing, while in our case we manufacture millions of contacted nanowires in a few min.

The exact growth mechanism of nw's on metals is not known. It may be semiconductor epitaxy on the metal, or semiconductor self-assembly on the metallic substrate, or either vapor-liquid-solid (VLS) or vapor-solid-solid (VSS) growth from substrate side or from the droplet side on the nanowire top. Fig. 3 showing GaAs nw on Au implies VLS or VSS growth mechanism with epitaxy on the metals. Fig. 1 and Fig. 2(b,c) showing GaAs nw growth on Au/Pd, Ag, Cu, Ni, Ga, Al and Ti implies VLS or VSS growth mechanism with thermal annealing. In the case of GaAs growth on Au, the droplet on the top of the nw is from the Au film, but in the case of other growth processes it may be either from Ga or from catalyst metal-substrate. These growth mechanisms need to be investigated further.

The growth method allows in situ fabrication of substrate-side metallic interfaces to the nanowires. Most probably metal-semiconductor contacts has quality of thermally annealed interfaces and at certain growth conditions it is possible to achieve epitaxial quality metal substrate – semiconductor nanowire interface. Possible applications include electrical, optical or superconducting contacts. The grown nanostructures are promising for future optoelectronics, electronics and quantum electronics devices. For instance, epitaxial quality contact shown in Fig.3, if achieved between semiconductor and superconductor could lead to applications in quantum electronics [82]. Contacts shown in Fig.1 and Fig.2(b,c) are presumably of quality of thermally annealed metal-semiconductor junction, which has been investigated by other groups finding improved electrical transport characteristics [83-86], thus promising applications in solar cells, transistors , diodes, electrically driven nanowires lasers, etc...

## Conclusions

GaAs nanowires were grown directly on various metal (Au, Au/Pd, Ag, Cu, Ni, Ga, Al and Ti) films by MOVPE. The GaAs nanowires were studied by SEM and TEM. It was concluded that the growth method allows in situ fabrication of substrate-side (bottom-side) metal contacts to the GaAs nanowires.

## Acknowledgments

This work was partially supported by the Academy of Finland (projects numbers 128445, 127280).  G.S. Thanks H. Lipsanen for supervision and valuable comments, Chief Executive Editor of APEX Akira Yamada for valuable critics, V.Ovchinnikov, A.Kemppinen and M.Meschke for the support with equipment, Yanling Ge for HRTEM image.

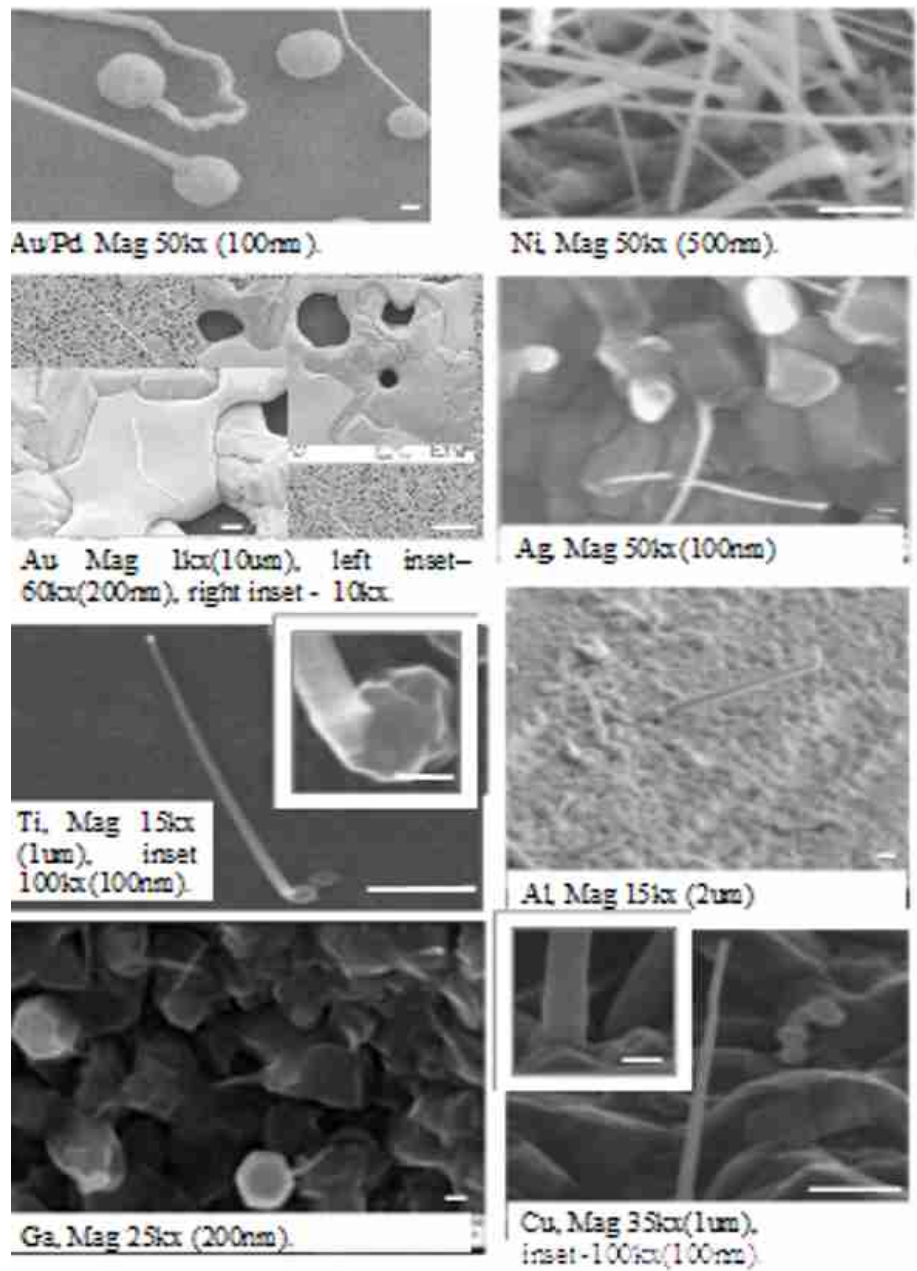

Fig.1. SEM images of GaAs nanowires and crystallites on metallized substrate. Described is the metal, magnification, Mag, and the scale of white bar on the right bottom corner (written in the brackets).

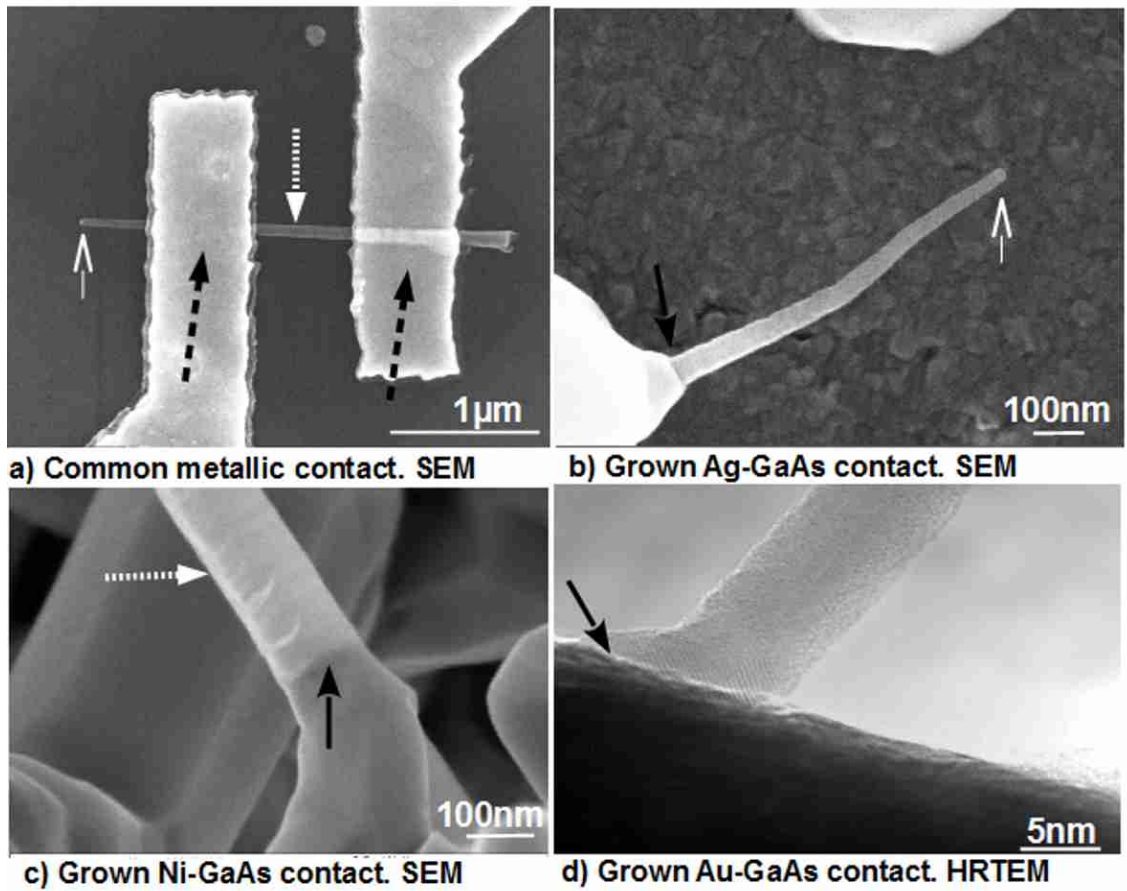

Fig. 2. Solid black arrows point to integer metallic contacts / junctions / interfaces (MSI) at the substrate end of nanowire formed by our approach during nanowire growth. Solid white concave shape arrows point to metal – semiconductor junction, which is common result in most catalysed ns growths. a): SEM image of contacts prepared by EBL. They are much larger than nanowires. #b-d: Novel metallic contact (b-Ag, c-Ni, d-Au) which is integer with GaAs nanostructure.

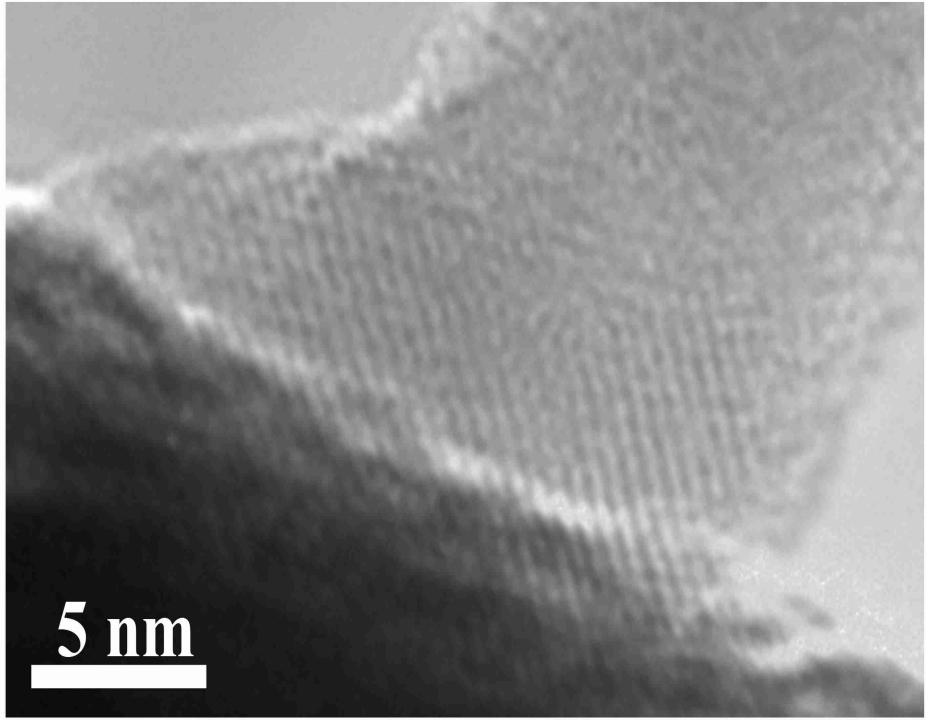

Fig. 3. HRTEM of Au - GaAs nanowire interface (magnified interface area from Fig.2d).